\newcommand{\printstyle}{reprint}
\newcommand{\halfwidth}{\columnwidth} 

\documentclass[aip,pop,floatfix,\printstyle ,nofootinbib]{revtex4-1}

\usepackage{color}
\newcommand{\addition}[1]{#1} 

\usepackage[caption=false]{subfig}
\usepackage{graphicx,ctable,booktabs,array}
\DeclareGraphicsExtensions{.pdf,.png,.jpg,.eps}
\usepackage{epsfig}
\usepackage{epstopdf}
\usepackage{amssymb}
\usepackage{amsmath}
\usepackage{amsfonts}
\usepackage{mathrsfs}
\usepackage{amsthm}
\usepackage{float}
\usepackage{amsbsy}
\usepackage{bm}
\usepackage{grffile}
\usepackage{hyperref}
\hypersetup{
    colorlinks = true,
    citecolor = blue,
    urlcolor = blue,
    linkcolor = blue,
}
\usepackage{courier}
\usepackage{upgreek}
\usepackage{xspace}

\newcommand{\etal}{\textit{et al}. }
\newcommand{\ie}{\textit{i}.\textit{e}. }
\newcommand{\eg}{\textit{e}.\textit{g}. } 

\newcommand{\Alfven}{Alfv\'{e}n }
\newcommand{\Alfvenic}{Alfv\'{e}nic }
\newcommand{\EGAE}{EP-GAE\xspace}

\newcommand{\va}{v_A}

\newcommand{\vb}{v_0}
\newcommand{\vc}{v_c}
\newcommand{\omegaci}{\omega_{ci}}
\newcommand{\omegacio}{\omega_{ci0}}
\newcommand{\omegator}{\omega_\phi}
\newcommand{\omegapol}{\omega_\theta}
\newcommand{\db}{\delta B}
\newcommand{\de}{\delta E}
\newcommand{\dbpar}{\db_\parallel}
\newcommand{\depar}{\de_\parallel}
\newcommand{\dbperp}{\db_\perp}
\newcommand{\kpar}{k_\parallel}
\newcommand{\kperp}{k_\perp}
\newcommand{\pphi}{p_\phi}
\newcommand{\pmin}{p_{\text{min}}}
\newcommand{\vinj}{\vb/\va}
\newcommand{\linj}{\lambda_0}
\newcommand{\dl}{\Delta\lambda}
\newcommand{\df}{\delta F}
\newcommand{\E}{\mathcal{E}}
\newcommand{\Epar}{\E_\parallel}
\newcommand{\Epres}{\Epar^{res}}
\newcommand{\Eperp}{\E_\perp}
\newcommand{\Jc}{\mathcal{J}}
\newcommand{\Jb}{J_\text{beam}}
\newcommand{\Jp}{J_\text{plasma}}
\newcommand{\Jnorm}{\Jb/\Jp}
\newcommand{\omeganorm}{\omega / \omega_{ci}}
\newcommand{\nbo}{n_b}
\newcommand{\neo}{n_e}
\newcommand{\nb}{\nbo / \neo}
\newcommand{\nv}{\nbo\vb / \neo\va}
\newcommand{\vpar}{v_\parallel}

\newcommand{\vdrift}{v_{\text{Dr}}}
\newcommand{\vperp}{v_\perp}
\newcommand{\lres}{\ell}
\newcommand{\fgrad}{\hat{\Pi}}
\newcommand{\J}[2]{\mathscr{J}_{#1}^{#2}}
\newcommand{\Jlm}{\J{\lres}{m}}
\newcommand{\Jlg}{\J{\lres}{G}}
\newcommand{\Bres}{\eta}
\newcommand{\oba}{\frac{\omega\alpha^2}{\omegaci}}

\newcommand{\avg}[1]{\left\langle #1 \right\rangle}
\renewcommand{\vec}[1]{\bm{#1}}

\newcommand{\abs}[1]{\left|#1\right|}
\renewcommand{\dot}{\cdot}
\newcommand{\cross}{\times}
\renewcommand{\div}{\nabla\dot}
\newcommand{\grad}{\nabla}
\newcommand{\curl}{\nabla\cross}
\newcommand{\defined}{\equiv}

\newcommand{\figref}[1]{Fig.\xspace\ref{#1}}
\renewcommand{\eqref}[1]{Eq.\xspace\ref{#1}}
\newcommand{\secref}[1]{Sec.\xspace\ref{#1}}
\newcommand{\citeref}[1]{Ref.\xspace\onlinecite{#1}}
\newcommand{\appref}[1]{Appendix\xspace\ref{#1}}

\newcommand{\code}[1]{\texttt{#1}\xspace}
\newcommand{\HYM}{\code{HYM}}
\newcommand{\TRANSP}{\code{TRANSP}}
\newcommand{\NUBEAM}{\code{NUBEAM}}
\newcommand{\NOVA}{\code{NOVA}}

\newcommand{\PPPL}{Princeton Plasma Physics Lab, Princeton, NJ 08543, USA}
\newcommand{\Princeton}{Department of Astrophysical Sciences, Princeton University, Princeton, NJ 08543, USA}

\begin{document}

\title{Energetic-particle-modified global \Alfven eigenmodes}
\author{J.B. Lestz}
\email{jlestz@pppl.gov}
\affiliation{\Princeton}
\affiliation{\PPPL}
\author{E.V. Belova} 
\affiliation{\PPPL}
\author{N.N. Gorelenkov}
\affiliation{\PPPL}
\date{\today}
\begin{abstract}
Fully self-consistent hybrid MHD/particle simulations reveal strong energetic particle modifications to sub-cyclotron global \Alfven eigenmodes (GAE) in low-aspect ratio, NSTX-like conditions. Key parameters defining the fast ion distribution function -- the normalized injection velocity $\vinj$ and central pitch -- are varied in order to study their influence on the characteristics of the excited modes. It is found that the frequency of the most unstable mode changes significantly and continuously with beam parameters, in accordance with the Doppler-shifted cyclotron resonances which drive the modes, and depending most substantially on $\vinj$. This unexpected result is present for both counter-propagating GAEs, which are routinely excited in NSTX, and high frequency co-GAEs, which have not been previously studied. Large changes in frequency without clear corresponding changes in mode structure are signatures of an energetic particle mode, referred to here as an energetic-particle-modified GAE (\EGAE). Additional simulations conducted for a fixed MHD equilibrium demonstrate that the GAE frequency shift cannot be explained by the equilibrium changes due to energetic particle effects. 
\end{abstract}
\maketitle
\section{Introduction}
\label{sec:Intro}
High frequency fluctuations identified as global \Alfven eigenmodes (GAE) and compressional \Alfven eigenmodes (CAE) are routinely excited in beam-heated, low aspect ratio tokamaks such as NSTX\cite{Gorelenkov2003NF,Gorelenkov2004POP,Crocker2013NF,McClements2017PPCF} and MAST\cite{Sharapov2014PP}. These modes are driven by the relatively large super-\Alfvenic ion population that results from the low toroidal field, though they have also been observed in DIII-D\cite{Heidbrink2006NF}. GAEs are ideal shear \Alfven MHD modes with frequencies lying just below minima of the \Alfven continuum, \eg $\omega_{GAE} \leq \left[\kpar(r)\va(r)\right]_\text{min}$. Their existence results from coupling to the magnetosonic mode, an equilibrium current, current density gradient, and finite $\omeganorm$ effects\cite{Appert1982PP,Mahajan1983PF,Mahajan1984PF,Li1987PF,DeAzevedo1991SP}. ``Nonconventional" GAEs may also be excited above a local maxima in the continuum through similar mechanisms\cite{Kolesnichenko2007POP}. Due to their separation from the continuum, some GAEs may avoid substantial continuum damping. Consequently, these modes can be driven unstable by the free energy in gradients in the energetic particle (EP) distribution. Instability requires energetic particles to resonate with the wave through the general Doppler-shifted cyclotron resonance $\omega - \kpar\vpar - \kperp\vdrift = \lres \omegaci$, with drive generated by the anisotropy in beam-like distributions. The cyclotron harmonic coefficient $\lres$ can be $-1$, $0$, or $1$ depending on the sign and magnitude of the Doppler shift. \addition{Note that here and for the rest of the paper, the ``Doppler shift" refers to the shift in the resonance due to a particle's parallel and drift motion, not the bulk rotation of the plasma.} 

GAEs were initially discovered in cylindrical plasmas\cite{Appert1982PP} (very large aspect ratio approximation), and later found to be stabilized by finite toroidicity effects\cite{Fu1989PF,VanDam1990FT}. These early works considered only the $\lres = 0$ Landau resonance. In NSTX, the beam injection velocity $\vb$ can be $3 - 6$ times larger than the \Alfven velocity $\va$, which can result in a Doppler shift large enough to satisfy the cyclotron resonance condition with $\lres=\pm 1$ for sufficiently large $\kpar$. Cyclotron resonance-driven GAEs which propagate against the direction of the plasma current (cntr-GAEs) have been studied extensively and are common in NSTX\cite{Gorelenkov2003NF,Gorelenkov2004POP,Fredrickson2012NF,Fredrickson2013POP,Crocker2013NF}. Higher frequency co-GAEs excited by the $\lres = -1$ resonance have not been studied before this work or observed experimentally. Moreover, most of the existing work on EP effects on high frequency \Alfven eigenmodes was focused on non-adiabatic beam effects, \ie beam contribution to the growth rate alone. The model used here includes all fast ion effects fully self-consistently, allowing excitation via cyclotron resonances, EP modifications to the equilibrium, and an adiabatic contribution to the GAE dispersion. 

A detailed study of GAEs properties is warranted because of their potential effects on plasma heating profiles. In
particular, in NSTX the presence of GAEs and CAEs has been linked to anomalously flat electron temperature profiles at high beam power\cite{Stutman2009PRL,Ren2017NF}, which limits fusion performance and could imperil future spherical tokamak development. The inferred electron diffusion profile needed to generate this flattening is not associated with any source of microturbulence seen in gyrokinetic simulations of the core region where gradients are absent\cite{Guttenfelder2013NF}. \addition{In these same discharges, the thermal ion diffusivity is close to neoclassical, and multiple diagnostics have ruled out large beam ion transport\cite{Stutman2009PRL}.} There are two previously proposed mechanisms in which GAEs and CAEs can modify the electron temperature profile. One involves the stochastization of electron orbits induced by the presence of many overlapping modes of sufficient amplitude\cite{Gorelenkov2010NF}. The other is an energy-channeling mechanism where a core-localized CAE or GAE mode converts to a kinetic \Alfven wave (KAW) at the \Alfven resonance location, which damps efficiently on electrons, effectively redirecting neutral beam power from the core to the edge\cite{Kolesnichenko2010PRL,Kolesnichenko2010NF,Belova2015PRL,Belova2017POP}. These mechanisms have been demonstrated numerically, though their quantitative predictions do not presently reproduce the experimental anomaly. Furthermore, GAEs are prone to frequency chirping in NSTX, which can modify the characteristics of fast ion transport and presents opportunities for validating nonlinear theories\cite{Duarte2017NF}. GAE ``avalanches'' -- sudden, broad spectrum, large amplitude bursts -- were also observed on NSTX, with implications for fast ion transport\cite{Fredrickson2012NF}. Further investigation of the character and properties of the sub-cyclotron \Alfven eigenmodes in simulations is motivated by their impact on the thermal plasma. 

Linear 3D hybrid simulations presented here demonstrate that the high frequency shear \Alfven waves excited in NSTX conditions can be strongly nonperturbative -- a fact that has not been recognized before. Consequently, this mode could be considered an energetic particle mode, or an energetic-particle-modified global \Alfven eigenmode (\EGAE). This is primarily concluded due to large changes in the frequency of the most unstable mode in proportion to the maximum energetic particle velocity without clear corresponding changes in the mode structure or location tracking the minimum of the \Alfven continuum. This behavior is pervasive for both co- and counter-propagating modes for all examined toroidal mode numbers, $\abs{n} = 4 - 16$. If the resonant value of $\vpar$ is proportional to the injection velocity $\vb$, then the large frequency changes can be qualitatively explained by the resonance condition. The most unstable mode frequency is determined to a large degree by features of the energetic particle population, in addition to properties of the thermal plasma -- a key signature of energetic particle modes (EPM)\cite{Heidbrink2008POP}. These may be the first example of EPM-type fluctuations that are excited at a significant fraction of the ion cyclotron frequency, typically $\omeganorm \approx 0.1 - 0.5$. The goal of this paper is to study the properties of unstable {\EGAE}s in simulations, in order to guide future theoretical studies of these modes and enable experimental tests of their distinguishing features. 

This paper is organized as follows. The hybrid model used to simulate the plasma is described in \secref{sec:Model}. The primary simulation results which this paper seeks to explain are detailed in \secref{sec:Freq-Dependence}. The relative importance of changes to the equilibrium versus changes to the fast ions in accounting for this effect is investigated in \secref{sec:EQvsEP}. The poloidal mode structure of the excited modes is shown for a range of EP energies in \secref{sec:Structure}, and the frequency of the most unstable mode for a wide variety of beam parameters is compared against the shear \Alfven dispersion relation. Lastly, the characteristics of the resonant particles are examined in \secref{sec:Resonance} as a function of the injection energy in order to clarify the role that the resonant wave-particle interaction plays in setting the frequency of the most unstable mode. A summary of the key results and discussion of implications for NSTX-U is given in \secref{sec:Discussion}.
\section{Hybrid Model Description}
\label{sec:Model}
To study these modes numerically, the hybrid MHD/particle code \HYM\cite{Belova1997JCP,Belova2000POP,Belova2003POP} is used. \HYM is an initial value code run in full 3D toroidal geometry. A single fluid MHD thermal plasma is coupled to energetic ions treated kinetically with a full orbit $\df$ scheme. Full orbit physics must be retained for the fast ions in order to study waves excited by cyclotron resonances. These two components interact via current coupling through the thermal plasma momentum equation
\begin{equation}
\label{eq:mom}
\rho \frac{d \vec{V}}{d t} = -\nabla P + (\vec{J} - \vec{J_b}) \cross \vec{B} - e n_b (\vec{E} - \eta \delta\vec{J}) + \upmu \Delta \vec{V}
\end{equation}
Where $\rho, \vec{V}, P$ are the thermal plasma mass density, fluid velocity, and pressure. The energetic particle (beam) density and current are $n_b$ and $\vec{J_b}$. The total plasma current is determined by $\mu_0\vec{J} = \curl \vec{B}$ while $\mu_0\delta\vec{J} = \curl \delta\vec{B}$ is the perturbed current. Nonideal MHD physics are introduced through the viscosity coefficient $\upmu$ and resistivity $\eta$. In addition to \eqref{eq:mom}, the thermal plasma evolves according to the following set of fluid equations 
\begin{subequations}
\begin{align}
\label{eq:fluids1}
\vec{E} &= - \vec{V} \cross \vec{B} + \eta \delta\vec{J} \\
\label{eq:fluids2}
\frac{\partial \vec{B}}{\partial t} &= - \curl \vec{E} \\ 
\label{eq:fluids3}
\frac{\partial \rho}{\partial t} &= - \div \left( \rho \vec{V} \right) \\ 
\label{eq:fluids4}
\frac{d}{dt}&\left( \frac{P}{\rho^\gamma} \right) = 0 
\end{align}
\label{eq:fluids}
\end{subequations}
In fully nonlinear simulations \addition{(such as those presented in \citeref{Belova2017POP})}, the pressure equation includes Ohmic and viscous heating in order to conserve the system's energy. These effects are neglected in the linearized simulations presented here, reducing to the adiabatic equation of state in \eqref{eq:fluids4} with $\gamma = 5/3$. The nonlinear system conserves total energy exactly\cite{Burby2017PPCF}. \addition{The fields are evolved on a cylindrical grid ($z,R,\phi$), while the particle quantities are computed on a Cartesian grid $(z,x,y)$ sharing the $z$ grid points. A second order accurate mapping between these two grids is defined by quadratic splines. In these simulations, the particle grid has dimensions $(120,51,51)$, with 500,000 particles used to represent the fast ions. The field grid is of size $(120,120,64)$ when simulating modes with toroidal mode numbers $\abs{n} < 8$, and a grid with dimensions $(120,96,128)$ is used to resolve the higher mode numbers. For specific cases, up to four times larger grid sizes and 20 million particles have been tested, which can result in slightly different growth rates but has no impact on frequency or mode structure. All simulations used a time step of $\Delta t = 0.05 \omegacio$ to evolve the vector potential $\vec{A}$, bulk momentum, density, pressure, and particle quantities. These fields are then used to update the remaining field quantities.}

The EP distribution is decomposed into an equilibrium and perturbed part, $F = F_0 + \df$. Each numerical particle has a weight $w = \df / P$ where \addition{$P$ is a function of integrals of motion used for particle loading $(dP/dt = 0)$.} These $\df$ particles representing the fast ions evolve according to the equations of motion in \eqref{eq:parts}
\begin{subequations}
\begin{align}
\label{eq:parts1}
\frac{d\vec{x}}{dt} &= \vec{v} \\
\label{eq:parts2}
\frac{d\vec{v}}{dt} &= \frac{q_i}{m_i}\left(\vec{E} - \eta \delta\vec{J} + \vec{v} \cross \vec{B}\right) \\ 
\label{eq:dwdt}
\frac{dw}{dt} &= -\left(\frac{F}{P} - w\right)\frac{d \ln F_0}{dt}
\end{align}
\label{eq:parts}
\end{subequations}
\addition{Since $w \propto \df$,} particles with large weights indicate regions of phase space with strong wave-particle interactions. \addition{The simulations included in this study are linear, meaning that the fluid equations described in \eqref{eq:fluids} omit nonlinearities in fluctuating quantities and the particle trajectories are unperturbed from their equilibrium paths. The equation for particle weights is also linearized, \ie the $w$ term is dropped from the RHS of \eqref{eq:dwdt}. Particle weights are used to calculate the $\delta n_b$ and $\delta\vec{J}$ terms which appear in \eqref{eq:mom}. }

The equilibrium fast ion distribution function is written as a function of the constants of motion $\E$, $\lambda$, and $\pphi$. The first, $\E = \frac{1}{2}m_i v^2$, is the particle's kinetic energy assuming no equilibrium electric field. Next, $\lambda = \mu B_0 / \E$ is a pitch angle parameter, where first order corrections in $\rho_{EP}/L_B$ to the magnetic moment $\mu$ are kept for improved conservation\cite{Belova2003POP}. This correction is more relevant in spherical tokamaks than conventional tokamaks since the fast ion Larmor radius can be a significant fraction of the minor radius. In addition, $\lambda$ can be regarded as a trapping parameter, since $\lambda < 1 - \epsilon$ corresponds to passing particles, and particles with $1 - \epsilon < \lambda < 1 + \epsilon$ are trapped, where $\epsilon = r/R_0$. Lastly, $\pphi = -q_i \psi + m_i R v_\phi$ is the canonical toroidal angular momentum, conserved due to the axisymmetric equilibria used in these simulations. \addition{In the previous expression, $\psi$ is the poloidal magnetic flux, and below $\psi_0$ is its on-axis value.} The distribution is assumed to be a product of nearly single variable distributions: $F_0 (v,\lambda,\pphi) = F_1 (v) F_2 (\lambda) F_3 (\pphi,v)$ defined by \eqref{eq:F0}
\begin{subequations}
\begin{align}
\label{eq:F1}
F_1(v) &= \frac{1}{v^3 + v_c^3} \quad \text{ for } v < \vb \\
\label{eq:F2}
F_2(\lambda) &= \exp\left(-\left(\lambda - \linj\right)^2 / \dl^2\right) \\ 
\label{eq:F3}
F_3\left(\pphi,v\right) &= \left(\frac{\pphi - \pmin}{m_i R_0 v - q_i \psi_0 - \pmin}\right)^\sigma \text{ for } \pphi > \pmin
\end{align}
\label{eq:F0}
\end{subequations}
The energy dependence, $F_1(v)$, is a slowing down function with injection velocity $\vb$ and critical velocity $\vc$ \addition{(also known as the ``crossover velocity'' \cite{Gaffey1976JPP})}. A beam-like distribution in pitch is used for $F_2(\lambda)$, centered around $\linj$ and width $\dl$. Characteristic profiles of beam density calculated by the global transport code \TRANSP\cite{Goldston1982JCP} and Monte Carlo fast ion module \NUBEAM\cite{Pankin2004CPC} motivate the polynomial form of $F_3(\pphi,v)$. A prompt-loss boundary condition at the last closed flux surface is imposed by requiring $\pphi > \pmin = -0.1\psi_0$. \HYM is capable of including the energetic particles self-consistently or ignoring them when solving for the equilibrium. Inclusion of the fast ions results in a modified Grad-Shafranov equation\cite{Belova2003POP}
\begin{equation}
\label{eq:gs}
\frac{\partial^2\psi}{\partial z^2} + R\frac{\partial}{\partial R}\left(\frac{1}{R}\frac{\partial\psi}{\partial R}\right) 
= -R^2 P^\prime - HH^\prime - GH^\prime + RJ_{b\phi}
\end{equation}
$G(R,z)$ is a poloidal stream function for the beam current, defined as $\vec{J_{b,\text{pol}}} = \grad G \times \grad\phi$. $H(\psi)$ is defined from $h(R,z) = H(\psi) + G(R,z)$ and $h(R,z)$ appears in the equilibrium field $\vec{B} = \grad\phi \cross \grad\psi + h\grad\phi$. The last two terms on the right hand side of \eqref{eq:gs} are the contributions from the fast ions, which can generate pressure anisotropy, an increased Shafranov shift, and more peaked current profiles. Although the beam density is small, $\nbo \ll \neo$, the current carried by the beam can nevertheless be comparable to the thermal plasma current due to the significant difference in energy between the fast ions and thermal particles. \addition{For example, the ratio of energy stored in the beam ions relative to the bulk thermal plasma can exceed 30\%.} Since this study focuses on changes to GAE frequencies as a function of the EP distribution, understanding the effects of the energetic particles on the equilibrium can not be ignored. Specifically, changes to the equilibrium may influence the mode frequencies since the \Alfven continuum is sensitive to the thermal plasma profiles. Hence, an accurate explanation of how the GAE frequencies change with energetic particle parameters requires proper accounting of the effects of the energetic particles on the equilibrium. Presently, thermal plasma rotation is not included in the equilibrium. 
\section{Frequency Dependence on Fast Ion Parameters}
\label{sec:Freq-Dependence}
\begin{figure*}[tb]
\includegraphics[width = .8\textwidth]{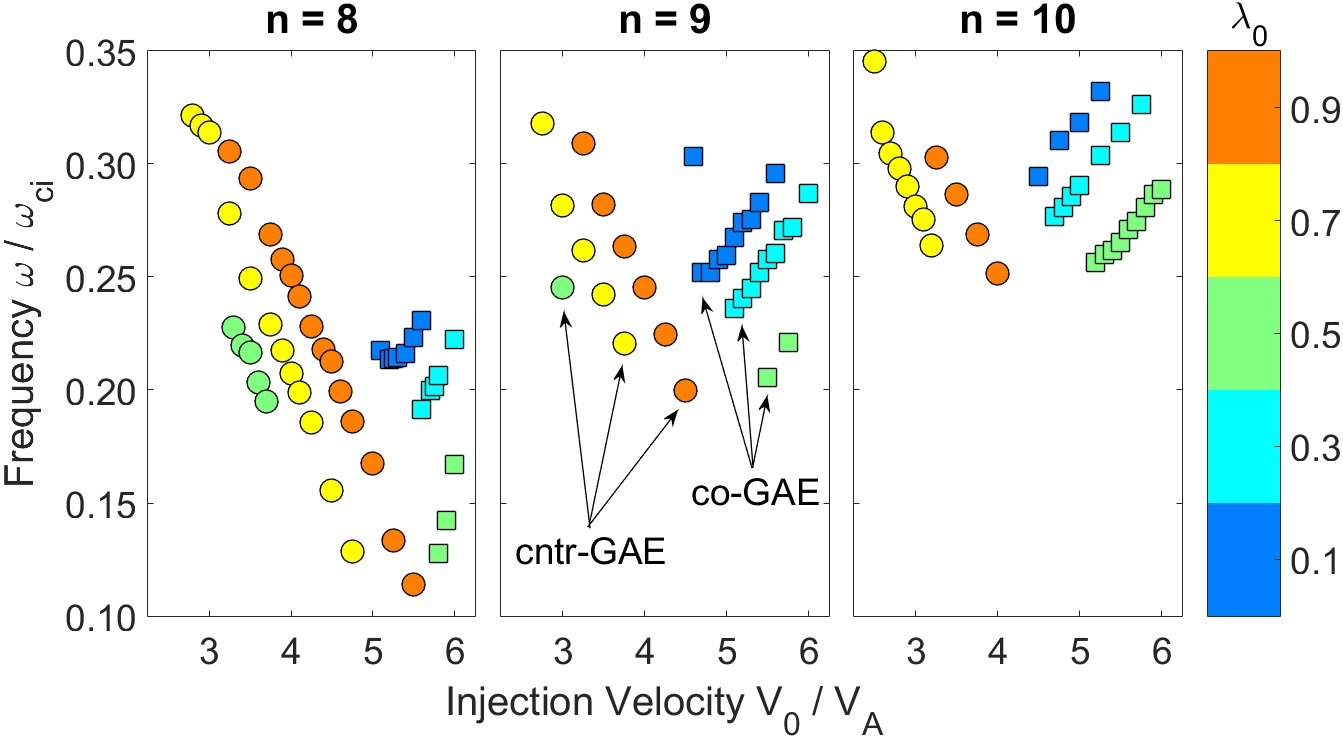}
\caption{Change in frequency for $\abs{n} = 8 - 10$ GAEs as a function of normalized injection velocity $\vinj$. Cntr-GAEs are marked by circles, and co-GAEs are marked by squares. Color denotes the central pitch $\linj$ of the EP distribution in each simulation.}
\label{fig:all_shifts}
\end{figure*}
This section describes results obtained from the self-consistent hybrid simulations, \eg those with an equilibrium that self-consistently includes fast ion effects. Since linear initial value simulations are conducted, only the mode with the largest growth rate can be seen. Consequently, the results in this section represent the properties of the most unstable mode in each simulation. A filter for a single toroidal harmonic is imposed on the simulation so that many distinct eigenmodes can be studied independently. 

Each of the simulations is based on the conditions of the well-analyzed NSTX H-mode discharge 141398\cite{Fredrickson2013POP,Crocker2013NF}, which has nominal experimental beam parameters of $\nb = 0.053$ and $\vinj = 4.9$, while $\linj = 0.7, \dl = 0.3, \vc = \vb/2,$ and $\sigma = 6$ are  chosen to reproduce the beam ion distribution function calculated by \NUBEAM. In ordinary NSTX operations, $\vinj = 3 - 6$ and $\linj \approx 0.5 - 0.7$. In this set of simulations, the normalized injection velocity $\vinj$ and the central pitch $\linj$ of the energetic particle distribution are varied in order to explore their effect on characteristics of the excited sub-cyclotron modes. Generally, unstable modes in the simulations are identified as GAEs instead of CAEs when $\dbperp \gg \dbpar$ near the plasma core. This identification is supported by previous cross validation between experiment, \HYM, and the \NOVA eigenmode solver. These efforts revealed good agreement between experimental measurements, GAEs found by \NOVA, and the shear-polarized modes excited in \HYM simulations\cite{Gorelenkov2003NF,Gorelenkov2004POP,Crocker2013NF,Crocker2017IAEA}. The modes identified as GAEs have linear growth rates ranging from $\gamma/\omegaci = 0.1 - 5\%$, with most around $1\%$ or less. Normalized instead to the mode frequency yields $\gamma/\omega = 1 - 20 \%$, with a few percent typical. 

\begin{figure*}[tb]
\includegraphics[width = .9\textwidth]{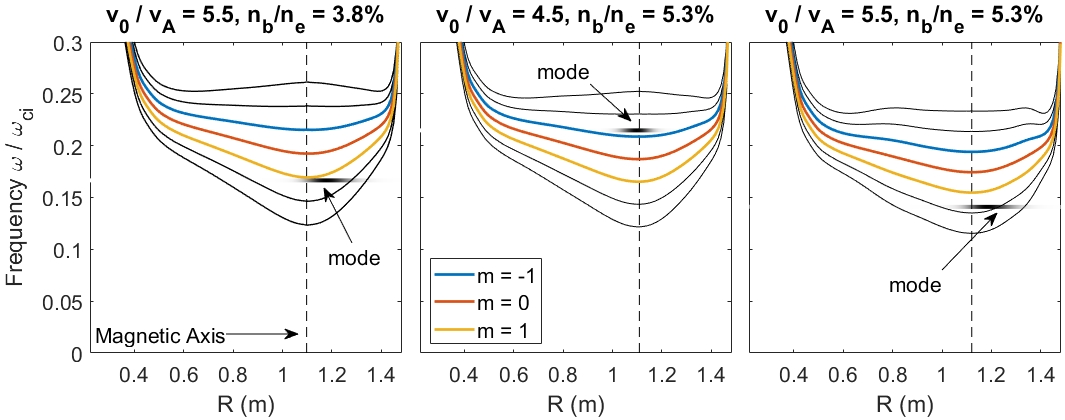}
\caption{\Alfven continuum for $n = 6$, including poloidal harmonics with $\abs{m} \leq 3$, for the self consistent equilibrium with different beam parameters -- left: $\vinj = 5.5$, $\nb = 3.8\%$, center: $\vinj = 4.5, \nb = 5.3\%$, right: $\vinj = 5.5, \nb = 5.3\%$. The thick horizontal lines mark the frequency and location of the mode excited in each simulation, where the darkness is proportional to the average amplitude of $\dbperp$. The $m = -1,0,1$ branches of the continuum are labeled for reference.}
\label{fig:continuum}
\end{figure*}

Unexpectedly, the frequency of the most unstable GAE for a single toroidal harmonic changes significantly as the energetic particle distribution is changed from one simulation to the next. The change in frequency is not usually accompanied by significant changes in the mode structure. Most notably, varying the injection velocity by a factor of two results in a factor of two change in the mode frequency. Since these modes are a non-negligible fraction of the cyclotron frequency ($\omeganorm \approx 0.1 - 0.5$), this can represent a dramatic change in frequency of hundreds of kilohertz. As GAEs are expected to have frequencies slightly below a minimum of the \Alfven continuum, such large changes in frequency with beam parameters clashes with their orthodox MHD description. In contrast, CAEs excited in similar simulations do not exhibit this same strong frequency dependence on fast ion parameters. Instead, the frequency of the most unstable CAE is nearly constant except for jumps in frequency at specific values of $\vinj$, which are also accompanied by a clear change in poloidal mode number\cite{Lestz2018sim}.

For sufficiently large beam injection velocities, GAEs propagating both with and against the direction of plasma current/beam injection are excited in the simulations. Analysis of the wave-particle interactions shows that co-GAEs and cntr-GAEs are driven by the Doppler-shifted cyclotron resonance with $\lres=-1$ and $\lres=1$, respectively. Counter-propagating GAEs are commonly observed in NSTX discharges while the co-propagating GAEs are yet to be detected. This is primarily due to geometric constraints of the neutral beam sources, since the co-GAEs are typically excited in the simulations when the energetic particle population has very low values of $\linj \lesssim 0.5$\cite{Lestz2018sim}, whereas the typical regime for NSTX is $\lambda \approx 0.5 - 0.7$. The additional beam sources on NSTX-U are more tangential and thus different beam mixtures could potentially excite modes propagating in either direction in future experiments, given sufficiently large $\vinj$.

For cntr-GAEs, the frequency of the most unstable mode decreases as injection velocity increases, whereas it increases for co-GAEs. \figref{fig:all_shifts} shows how the frequency changes with the normalized injection velocity $\vinj$ for each toroidal mode number $\abs{n} = 8 - 10$, where both co- and counter-propagating GAEs are excited in this set of simulations. Each point on the figure represents an individual simulation conducted with the energetic particle distribution from \eqref{eq:F0} parametrized by values of $\left( \vinj, \linj \right)$ in a 2D beam ion parameter scan. For each distribution, the equilibrium is re-calculated to self-consistently capture the EP effects on the thermal plasma profiles. It is clear that the frequency of the most unstable mode in each simulation depends linearly on the injection velocity, except for some outliers near marginal stability. The central pitch $\linj$ of the distribution also impacts the frequency, though this effect is not as pronounced. Especially noteworthy is the continuous nature of the change in frequency with injection velocity. 

Even at the smallest investigated increments of $\Delta\vinj = 0.1$, the change in frequency remains proportional to the change in injection velocity. This suggests the existence of either a continuum of modes which are being excited or very densely packed discrete eigenmodes. In the case of discrete eigenfrequencies, one would expect to see a discontinuous ``staircase" pattern in the frequency of the most unstable mode as a function of the injection velocity; a single discrete eigenmode with constant frequency would be the most unstable for some range of $\vinj$, with a jump to a new frequency when a different discrete mode becomes more unstable for the next velocity range. However, this is not what is observed, at least to the resolution of $\Delta\vinj = 0.1$. Overall, GAEs propagating with or against the plasma current exhibit a change in frequency proportional to the change in the normalized injection velocity of the energetic particles. The direction of this change matches the sign of $\kpar$, implicating the Doppler shift in the resonance condition as the likely explanation.

\addition{Moreover, these modes are global eigenmodes in the sense that the fluctuations oscillate at the same frequency at all points in space, and that the mode structure is converged at long times (once the mode has grown long enough to dominate the initial random perturbations).} Comparing the location of these modes relative to the \Alfven continuum can also help elucidate the character of these modes. Since these modes have been identified as GAEs in previous experimental and numerical analysis, one would expect them to be radially localized near a local minimum of the continuum with frequency near that value. For example, previous \HYM simulations of a separate NSTX discharge with smaller $\nb$ demonstrated excitation of a GAE with the expected characteristics, in particular with a frequency just below a minimum of the \Alfven continuum\cite{Gorelenkov2003NF,Gorelenkov2004POP}. If instead the modes substantially intersect the continuum, strong continuum damping would make their excitation unlikely, or suggest that they may not be shear \Alfven eigenmodes at all. The continuum is calculated using the $q(r)$ and $n(r)$ profiles from the self-consistently calculated equilibrium for three separate cases, and shown in \figref{fig:continuum}. The left-most case has $\vinj = 5.5, \,nb = 3.8\%$, and the mode peaks quite close to an on-axis minimum of the continuum. In the middle figure, $\vinj = 4.5,\,\nb = 5.3\%$, and the GAE actually occurs above the minimum, but nonetheless avoids intersecting the continuum due to its limited radial extent. The right-most case is $\vinj = 5.5,\, \nb = 5.3\%$, and moderately overlaps the continuum. These examples demonstrate that as the relative fast ion pressure becomes larger, either through increased density or energy, the modes can depart from their textbook description. 

A limitation of this analysis is that kinetic corrections to the MHD continuum could become important for an accurate comparison in this regime. For instance, Kuvshinov has shown that in a single fluid Hall MHD model, the kinetic corrections to the shear \Alfven dispersion due to finite Larmor radius effects is $n_b \kperp^2\rho_\perp^2 / n_e (1 + \kperp^2\rho_\perp^2)$, which is equivalent to a Pad{\'e} approximation to the full ion-kinetic response\cite{Kuvshinov1994PPCF}. Near peak beam density, $n_b/n_e$ can approach $20\%$ in these simulations, and large fast ion energies can yield $\kperp\rho_\perp \approx 2$, which yields a roughly $15\%$ correction from this term. Developing a model of the continuous spectrum including fast ions self-consistently would make this comparison more definitive, but is beyond the scope of this work, as it represents a quite substantial enterprise itself. 
\section{Equilibrium vs Fast Ion Effects}
\label{sec:EQvsEP}
The purpose of this paper is to determine numerically if these large changes in frequency (as large as $20 - 50\%$, or $100 - 500$ kHz) can be explained by energetic particle effects, or if they can be interpreted some other way. Since the preceding results were from simulations which included EP effects self-consistently in the equilibrium, one possible explanation is that increasing the beam energy is modifying the equilibrium (and \Alfven continuum), indirectly changing the characteristic GAE frequency. While $\nb$  is small (of order $5\%$) in these simulations, the fast ion current can be comparable to the thermal plasma current due to large beam energies. Previous work has demonstrated the substantial effects that the beam contribution can have on the equilibrium\cite{Belova2003POP}. Moreover, there is recent work showing that the inclusion of alpha particles can significantly deform the \Alfven continuum\cite{Slaby2016POP}. It is important to investigate if these changes in frequency can be attributed to changes in the self-consistent equilibrium or changes in the fast particles driving the mode, independent of the equilibrium. The latter would be typical of nonperturbative energetic particle modes while the former would fit well with an MHD description of GAEs. 
\begin{figure}[tb]
\subfloat{\includegraphics[width = \halfwidth]{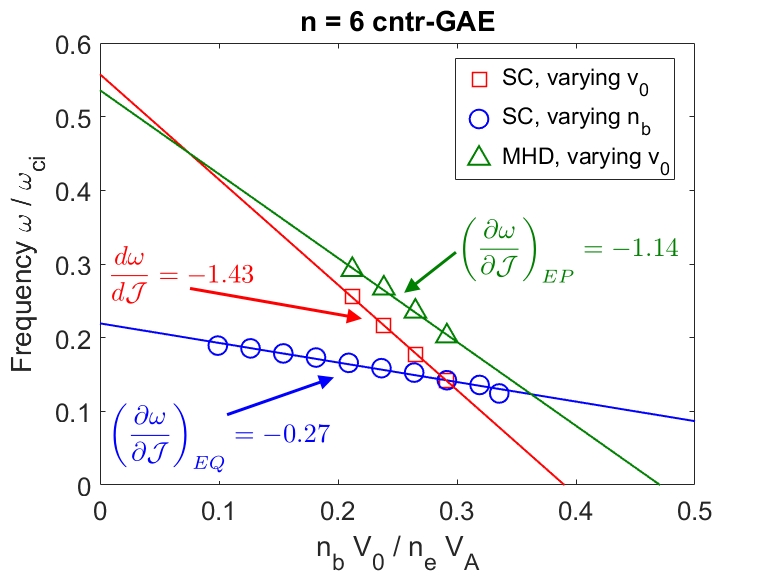}} \newline
\subfloat{\includegraphics[width = \halfwidth]{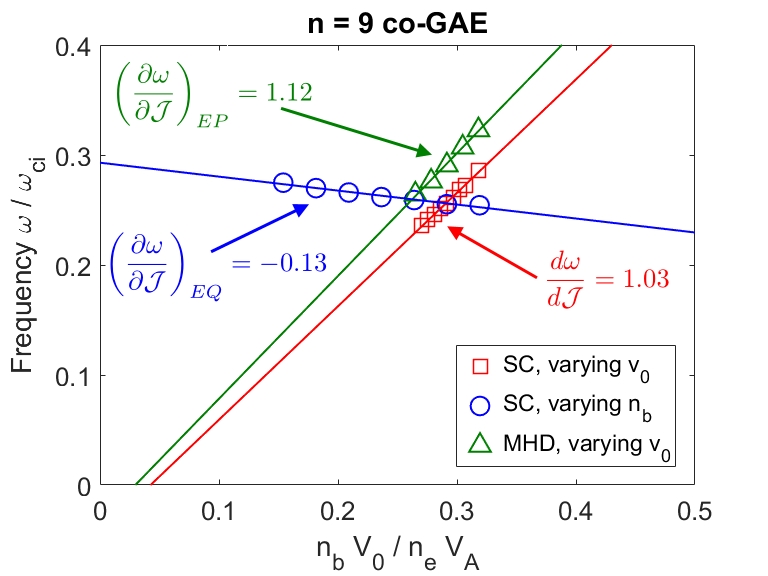}}
\caption{Frequency changes of modes as $\Jc = \nv \propto \Jnorm$ is varied under different conditions. (Red) Equilibrium includes EP self-consistently (``SC"); injection velocity $\vinj$ is varied while beam density $\nb$ is constant. (Blue) SC equilibrium; $\nb$ is varied, $\vinj$ is constant. (Green) Equilibrium determined without EP contributions (``MHD-only"); $\vinj$ is varied, $\nb$ is fixed. Top: counter-propagating $n = 6$ mode. Bottom: co-propagating $n = 9$ mode.}
\label{fig:sc_mhd}
\end{figure}
\subsection{Equilibrium Effects}
\label{sec:EQeffect}
In order to distinguish between these competing interpretations, these simulations were first reproduced at decreased EP density, since this decreases the ratio of the beam current to thermal plasma current, which is the key parameter controlling the impact of EP effects on the equilibrium profiles. These additional simulations are conducted for representative examples of both counter- and co-propagating GAEs. In the former case, an $n = 6$ mode driven by a beam distribution parametrized by $\vinj = 5.5, \linj = 0.7$ is studied, and for the latter, an $n = 9$ mode driven by a $\vinj = 5.5, \linj = 0.3$ distribution is selected. By varying $\nb$ with fixed $\vinj$ and combining with the previous simulation results which were conducted for constant $\nb$ and varying $\vinj$, the frequencies can be plotted against $\Jc \equiv \nv \propto \Jnorm$. If the frequency depends on this parameter in the same way in both sets of simulations, then it can be concluded that the large changes in frequency of the GAEs seen in the simulation are due to the EP-related changes to the equilibrium. 

The results of this comparison are shown in the top plot of \figref{fig:sc_mhd} for the cntr-GAE modes and the bottom for the co-GAEs. The red squares are simulations with fixed beam density and differing injection velocity (same conditions as those shown in \figref{fig:all_shifts}) whereas the blue circles show simulations where the EP distributions share a single value of $\vinj$ and have varying $\nb$. For both co- and cntr-GAEs, increasing beam density results in a modest decrease in mode frequency. This likely reflects changes in the equilibrium, and is supported by work done by Slaby \etal which found that the continuum frequencies are decreased in the presence of increased alpha particle pressure\cite{Slaby2016POP}. Also apparent in this comparison is that the mode has a different stability threshold in $\Jc$ depending on if $\Jc$ is decreased through $\nb$ or $\vinj$, as the mode can still exist for small $\Jc$ provided that $\vinj$ is sufficiently large. The mode frequency exhibits a linear dependence on EP density, with the slope for the two modes studied differing by a factor of two. The change in frequency due to this effect is less than $20\%$ of the magnitude of the change due to changing beam energy at constant beam density. Moreover, it has the opposite sign of that seen in the first set of simulations for the co-GAEs, which increase in frequency as $\vinj$ increases. These results demonstrate that changes to the equilibrium, proportional to $\Jnorm$, are not the primary cause of the large changes in frequency.
\subsection{Fast Ion Effects}
\label{sec:EPeffect}
Since the previous results suggest that the frequency changes can not be an equilibrium effect alone, the direct effects of the energetic particles should be isolated from the changes in the equilibrium. To do this, complementary simulations are conducted where the equilibrium is no longer calculated self-consistently to include the beam contribution. Instead, the equilibrium is solved for considering only the effects of the thermal plasma. This ``MHD-only" equilibrium is calculated with the same total current as the self-consistent one, and the plasma pressure is set to be comparable to the total thermal and beam pressure. These simulations will serve as a definitive test of the effects of the different energetic particle parameters on the excited mode frequency and structure for a single, fixed equilibrium. 

The simulations are repeated for the same $n = 6$ counter- and $n = 9$ co-propagating GAEs as introduced in \secref{sec:Freq-Dependence}. The results correspond to the green triangles on \figref{fig:sc_mhd}. The simulations with the fixed ``MHD-only" equilibrium and changing beam velocity reproduce the trend and approximate magnitude of the frequency shifts observed in simulations with the self-consistent equilibria (labeled ``SC" on the figure) for both the $n=6$ cntr-GAEs and $n=9$ co-GAEs. In order to distinguish between the various frequency dependencies, the following conventions are adopted for the different types of simulations conducted. $d\omega/d\Jc$ is the slope of the most unstable mode frequency with respect to $\Jc$ for simulations conducted with self-consistent equilibria and varying $\vinj$, which are the red squares on \figref{fig:sc_mhd}. These simulations represent the total frequency dependence on $\Jc$ since the changes to $\vinj$ alter both the equilibrium profiles and the location of resonant particles in phase space (detailed in \secref{sec:Resonance}). Changes in frequency in simulations with self-consistent equilibria with varying $\nb$ only, the blue circles, are purely due to changes in the equilibrium, so that slope is labeled as $\left(\partial\omega/\partial\Jc\right)_{EQ}$. Varying $\vinj$ for a fixed MHD-only equilibrium is a pure energetic particle effect on the frequency, associated with $\left(\partial\omega/\partial\Jc\right)_{EP}$ and shown as the green triangles. The effects on the GAE frequencies due to equilibrium and energetic particle effects appear to be nearly linear, succinctly stated in \eqref{eq:dflin}, which is accurate to within $5\%$ for the two cases studied in \figref{fig:sc_mhd}. This further supports that there are two independent factors determining the GAE frequency, and that the nonperturbative energetic particle influence on the mode dominates over the effects due to EP-induced changes to the equilibrium. 
\begin{equation}
\label{eq:dflin}
\frac{d \omega}{d \Jc} \approx 
\left(\frac{\partial\omega}{\partial \Jc}\right)_\text{EQ} + 
\left(\frac{\partial\omega}{\partial \Jc}\right)_\text{EP} = 
\neo\va
\left[
\frac{1}{\vb}\frac{\partial\omega}{\partial \nbo} + 
\frac{1}{\nbo}\frac{\partial\omega}{\partial\vb}
\right]
\end{equation}
For completeness, a final set of ``MHD-only" simulations were conducted where the beam energy is fixed and the beam density is varied. The changes in frequency due to varying this parameter are much smaller than any other, though they imply a negative partial derivative for both types of modes, similar to the SC EQ effect. This effect is labeled NR for non-resonant since it results from changes to the energetic particles, but not how they resonantly interact with the mode. It can be attributed to the small change of the continuum frequencies due to the change in total density when $\nbo$ is changed. For small $\nb$, this can be estimated as $\left.\partial\omega/\partial\Jc\right|_{\vinj} = -(\va/2v_0)\left(\kpar B_0 / \sqrt{n_e}\right)$ which evaluates to a slope of approximately $-0.02$ for the cntr-GAE case and $-0.03$ for the co-GAE case, which are of the right magnitude to explain the effect shown in the figure, and also very close to the less than $5\%$ discrepancy in \eqref{eq:dflin}. The relative magnitudes of these different effects are summarized in \eqref{eq:dftot}.  
\begin{equation}
\label{eq:dftot}
\Delta\omega \approx 
\left(\Delta\omega\right)_\text{EP} \gg 
\left(\Delta\omega\right)_\text{EQ} \gg 
\left(\Delta\omega\right)_\text{NR}
\end{equation} 
\section{Mode Structure and Dispersion}
\label{sec:Structure}
In order to determine if these are ideal MHD eigenmodes or strongly energetic-particle-modified modes such as EPMs, inspection of the mode structure is necessary. If MHD modes, one would expect that changes in frequency would be associated with some qualitative change in mode structure, such as the presence of different poloidal or radial harmonics, marking a new eigenmode. Conversely, in a nonperturbative energetic particle regime, the mode structure can be preserved even as the frequency changes significantly, such as in the theory and observation of chirping modes\cite{Fredrickson2006POP,Podesta2012NF,Duarte2017PPCF} or in the case of fishbones\cite{McGuire1983PRL,Chen1984PRL}. In these simulations of GAEs, the mode structure is frequently qualitatively unaffected by the large changes in frequency which accompany changes in the normalized EP beam energy. Quantitative changes are typically subtle, including slight changes in radial location, mode width, or elongation. \addition{A key difference between chirping modes, fishbones, and the GAEs studied here is that the first two fundamentally involve nonlinear physics, whereas the latter is a linear mode with nonperturbative EP modifications.}

\begin{figure}[t]
\subfloat[Poloidal structure at a single toroidal angle, slice taken at angle shown by radial line in (b).]{\includegraphics[width = \halfwidth]{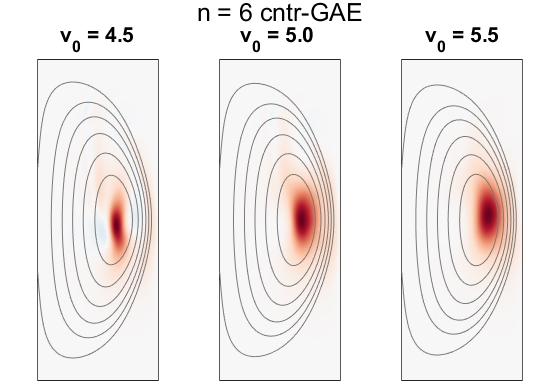}} \newline
\subfloat[Toroidal structure at midplane. Circles indicate the the last closed flux surface and magnetic axis.]{\includegraphics[width = \halfwidth]{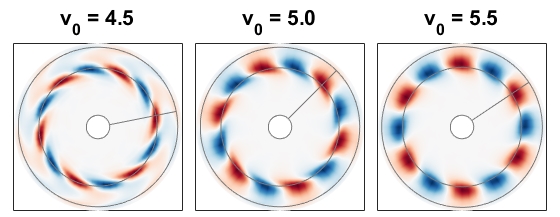}} \newline
\subfloat[Fourier amplitude of generalized poloidal harmonics along the $\vartheta = \nabla\psi \cross \nabla\phi$ direction, summed over all toroidal angles.]{\includegraphics[width = \halfwidth]{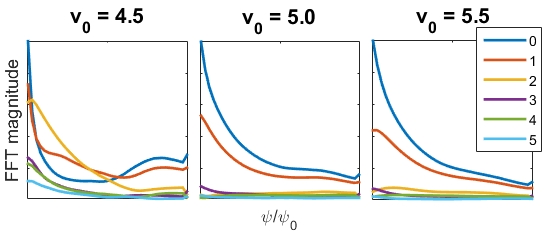}} 
\caption{Mode structure of $n = 6$ cntr-GAE excited by EP with $\linj = 0.7$ and $\vinj = 4.5, 5.0, 5.5$ in self-consistent simulations, with frequencies $\omeganorm = 0.214, 0.178, 0.141$. The fluctuation shown is $\dbperp$ in the $\nabla R \cross \vec{B_0}$ direction.}
\label{fig:cntr_struct}
\end{figure}
\begin{figure}[t]
\subfloat[Poloidal structure at a single toroidal angle, slice taken at angle shown by radial line in (b).]{\includegraphics[width = \halfwidth]{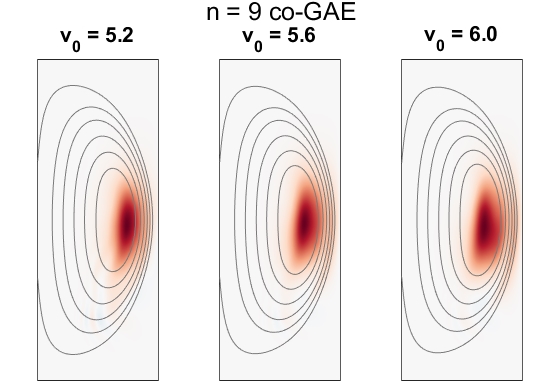}} \newline
\subfloat[Toroidal structure at midplane. Circles indicate the the last closed flux surface and magnetic axis.]{\includegraphics[width = \halfwidth]{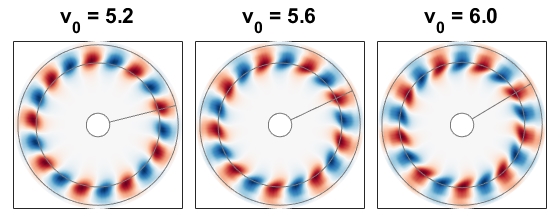}} \newline
\subfloat[Fourier amplitude of generalized poloidal harmonics along the $\vartheta = \nabla\psi \cross \nabla\phi$ direction, summed over all toroidal angles.]{\includegraphics[width = \halfwidth]{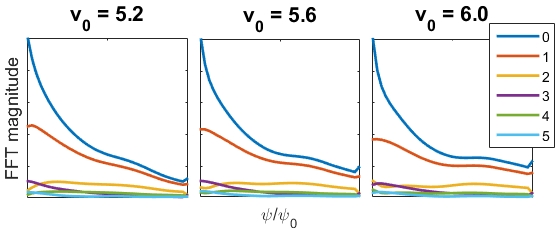}} 
\caption{Mode structure of $n = 9$ co-GAE excited by EP with $\linj = 0.3$ and $\vinj = 5.2, 5.6, 6.0$ in self-consistent simulations, with frequencies $\omeganorm = 0.239, 0.264, 0.289$. The fluctuation shown is $\dbperp$ in the $\nabla R \cross \vec{B_0}$ direction.}
\label{fig:co_struct}
\end{figure}

This endeavor is complicated by the fact that the GAEs, the counter-propagating modes especially, may interact with the continuum and excite a kinetic \Alfven wave, inferred through the presence of a well-localized $\depar$ fluctuation on the high field side and coincident short-scale modulation of the $\dbperp$ mode structure near this region. The coupling of the KAW with the compressional mode in \HYM simulations was studied in depth in a recent publication\cite{Belova2017POP}, which identified key signatures of the KAW in the simulation which can also be leveraged in the case of the GAEs. Some of the more dramatic changes in mode structure can be attributed to gradual suppression or excitation of KAW features, which has dominant $\dbperp$ polarization just as the GAEs do. This can be subjectively distinguished from the GAE mode structure since the KAW has a characteristic ``tilted" structure near the \Alfven resonance location whereas the GAE is usually concentrated between the axis and mid-radius, often towards the low-field side. \figref{fig:cntr_struct} shows how the mode structure evolves as a function of $\vinj$ for the $n = 6$ cntr-GAE in fully self-consistent simulations. Visually, the structure could be assigned a poloidal mode number of $m = 0$ or $m = 1/2$ since it has a single peak. Fourier decomposition in the generalized poloidal direction $(\vartheta = \nabla\psi \cross \nabla\phi)$ yields the same answer, some mix of $m = 0$ and $m = 1$. From $\vinj = 4.5$ (first column) to $\vinj = 5.5$ (last column), the frequency changes by $34\%$, or about 175 kHz, yet no new poloidal or radial harmonic emerges. Qualitatively, the structure becomes broader as $\vinj$ increases, and also gradually shifts towards the low field side, as can be seen in the midplane slices. 

For co-GAEs, there is even less change. Generally, the co-GAE mode structure is more broad radially and more elongated than the cntr-GAE structure. The poloidal structure of the co-GAEs looks very similar when excited by energetic particles with $\vinj = 5.2 - 6.0$, as shown in \figref{fig:co_struct}. Again, Fourier decomposition yields $m = 0 - 1$, matching visual intuition, and remaining unchanged as $\vinj$ is varied. For the case shown, the frequency changes by more than $20\%$, equivalent to 150 kHz. In contrast to the cntr-GAE, the co-GAEs migrate slightly towards the high field side for larger EP energies. Similar to the cntr-GAEs, this constancy of the mode structure despite large changes in frequency would be very atypical of MHD eigenmodes. Since these modes are $m = 0$ or 1 with $n = 9$, the approximation $\kpar \approx k_\phi = n/R$ is justified. Hence, this change in mode location to lower $R$ tends to increase $\kpar$. Furthermore, $\va$ has its minimum near the magnetic axis, so the local \Alfven speed can also change due to shifts in the mode location. It is then possible that a change in mode location could occur such that the frequency changes while conserving $\omega \approx \kpar \va$ without changing the mode numbers. However, this would necessarily move the mode away from an extremum in the \Alfven continuum (if it were originally near one when excited by lower $\vinj$), leading it to intersect the \Alfven continuum, which typically results in strong damping. This is essentially what was observed in the ``MHD-only" simulations and shown in \figref{fig:continuum}. 
\begin{figure}[tb]
\subfloat{\includegraphics[width = \halfwidth]{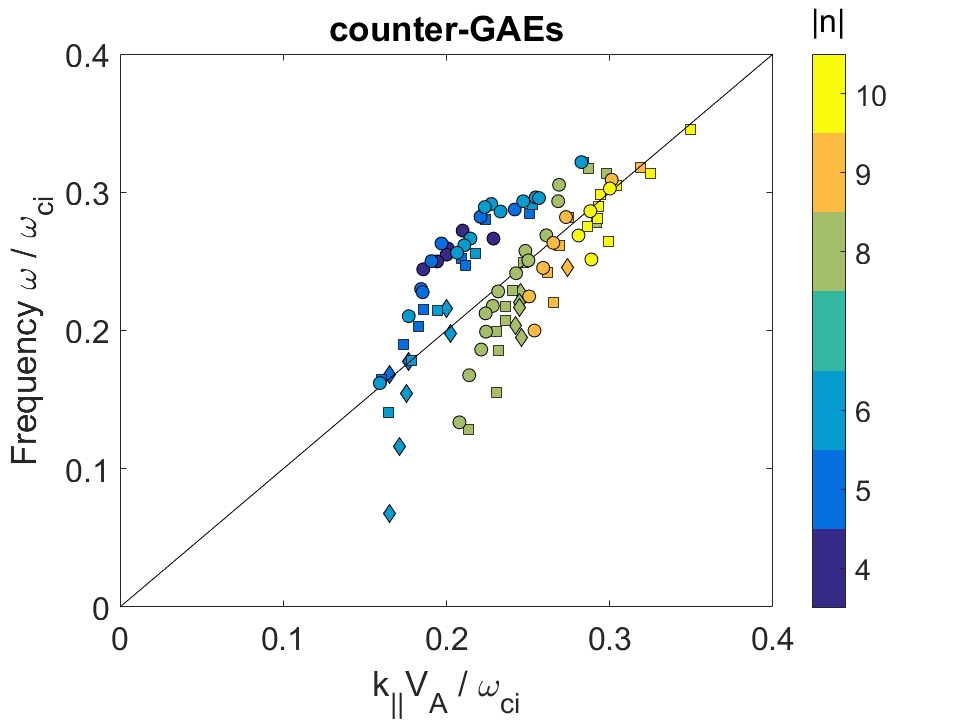}} \newline
\subfloat{\includegraphics[width = \halfwidth]{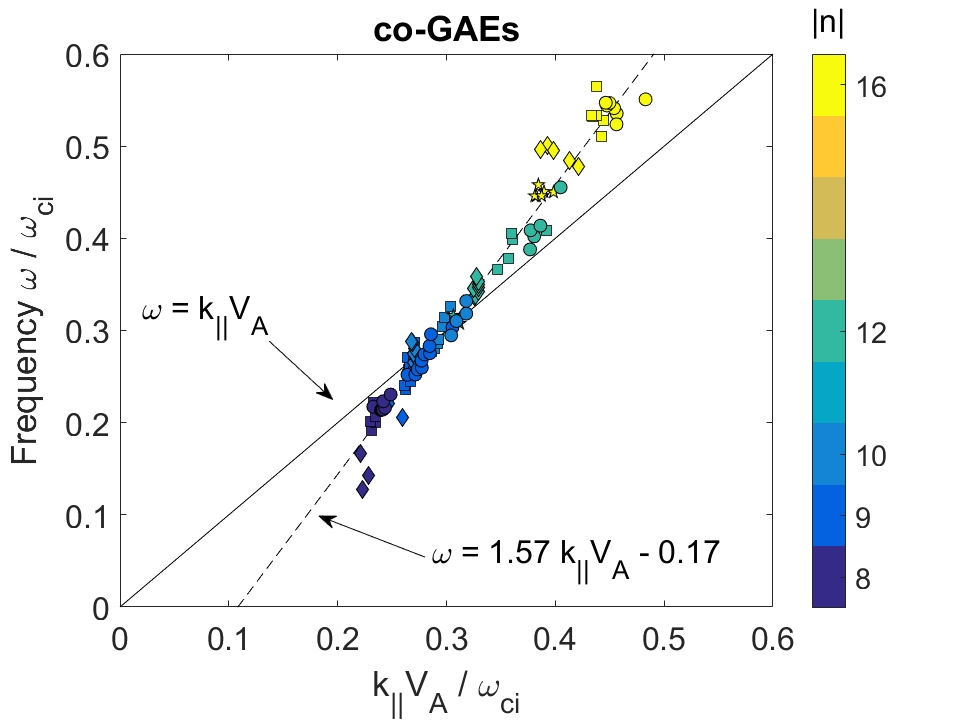}}
\caption{Comparison of mode frequencies to shear \Alfven dispersion with $\kpar$ and $\va$ evaluated at the peak mode location. Solid line indicates $\omega = \kpar\va$, dashed line indicates linear fit to simulation data. Color: toroidal mode number of the simulated mode. Top: cntr-GAEs. Bottom: co-GAEs.}
\label{fig:dispersion}
\end{figure}
Since counter-propagating \Alfven eigenmodes with shear polarization have typically been identified as perturbative GAEs in NSTX plasmas both experimentally\cite{Gorelenkov2003NF,Crocker2013NF} and in simulations\cite{Gorelenkov2004POP}, it is necessary to determine if their frequencies lie close to the shear \Alfven dispersion, $\omega_A = \kpar\va$, or if they deviate significantly due to the large frequency changes with beam parameters. While perturbative GAEs should have frequencies shifted somewhat below the \Alfven frequency, the difference should be small, \eg $\lesssim 10\%$ and often much less\cite{Appert1982PP}. For accuracy, the dispersion relation should be evaluated at the mode location. Calculating $\va$ at the mode location is only nontrivial due to the mode structure being broad, though this is easily solved by defining the mode location to be the $\db^2$ weighted average of $R$. The parallel wave number is less well defined. In a large aspect ratio tokamak, it is accurately represented by the familiar formula $\kpar = (n - m/q)/R$. However, this is only valid for $\epsilon = r/R \ll 1$ and requires $m$ to be well defined. In contrast, these simulations are carried out at the low aspect ratio of NSTX, where $\epsilon \approx 3/4$, and there is often no clear poloidal harmonic present in the mode structure, as discussed in section \ref{sec:Structure}. For high $n$ numbers, the approximation $\kpar \approx k_\phi$ becomes more reliable since typically $nq > m$ for the modes excited in the simulations. However, this is a poor approximation for the cntr-GAEs which may have, for instance, $n = 4$ and $m = 2 - 4$. As an alternative, the most literal interpretation of $\kpar$ is used, that is the peak in the Fourier spectrum of the fluctuation when projected onto the background field lines near the mode location with a field-line following code. This method is sufficient to determine if the mode frequencies are at least ``near" the shear \Alfven frequency, as in \figref{fig:dispersion}. 

For both counter- and co-propagating modes, there is a clear correlation between the frequency of the modes and the shear \Alfven dispersion, as expected for GAEs. However, the cntr-GAEs show significant deviation from this relation for low $\abs{n}$ modes, while the co-GAEs show a steeper than expected slope. The co-GAEs are well fit by the relation $\omega = 1.57\kpar\va - 0.17$. The deviations from the shear \Alfven dispersion are not explained at this time. A complete explanation likely requires modification of the GAE dispersion to include beam contributions to the eigenequation nonperturbatively, as well as coupling to the compressional mode. In order to remain consistent with the simulation results, the modification must at least include a term proportional to $\kpar\vb$. One route to pursue would be to build upon the theory developed by Berk \etal for reverse-shear \Alfven eigenmodes (RSAE) which employs energetic particle effects to localize the eigenmode near local extrema in the \Alfven continuum\cite{Berk2001PRL,Breizman2003PP}. In particular, Eq. 5 of \citeref{Berk2001PRL} includes terms proportional to $\avg{n_h}$ and $\kpar\avg{J_{\parallel h}}$ which could help explain the results in \figref{fig:sc_mhd}. The derivation of an accurate dispersion for the \EGAE is left for future work. 
\section{Resonant Particles}
\label{sec:Resonance} 
Ultimately, the resonance condition is determined to be responsible for key properties of these modes. Investigation of the properties of the resonant particles identified in the simulation with explanations supported by analytic theory can shed light on the origins of the unusual features of these modes. 
\begin{figure}[tb]
\subfloat{\includegraphics[width = \halfwidth]{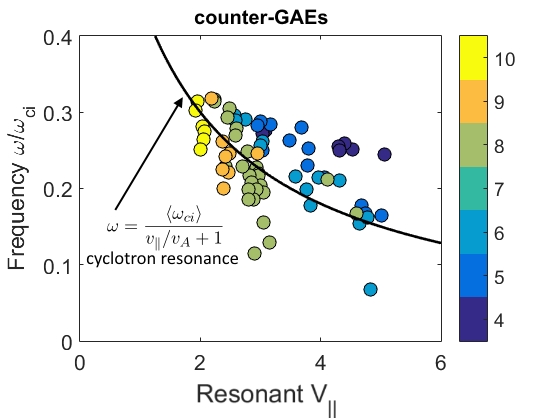}} \newline
\subfloat{\includegraphics[width = \halfwidth]{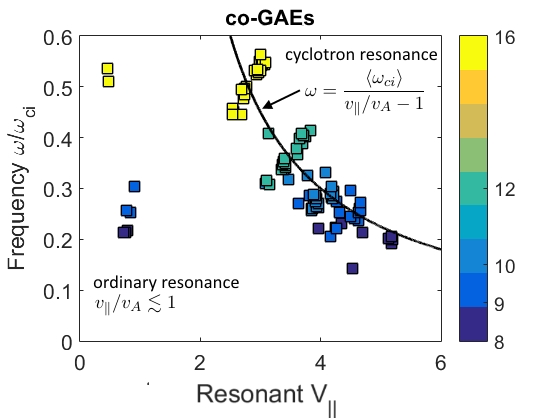}}
\caption{Frequency and approximate $\vpar$ of resonant particles with largest weights. Solid line is the expression from \eqref{eq:res_vpar} required by the dispersion and resonance condition, assuming $s = 0$. \addition{Frequency and velocity normalized by on-axis values of $\omegaci$ and $\va$.} Top: cntr-GAEs. Bottom: co-GAEs.}
\label{fig:resonance}
\end{figure}
\subsection{Influence of Resonance Condition}
\label{sec:ResA}
Since a $\df$ scheme is employed, the particle weights can reveal information about resonant particles. The weights will evolve according to \eqref{eq:dwdt}. Hence weights with large magnitudes correspond to regions of phase space with large changes in the distribution function, \eg particles which interact strongly with the waves. Particles can resonate with the wave through the general Doppler-shifted cyclotron resonance,
\begin{equation}
\label{eq:res_one}
\omega - \avg{\kpar\vpar} - \avg{\kperp\vdrift} \approx \omega - \avg{\kpar\vpar} = \lres\avg{\omegaci}
\end{equation} 
\addition{On the right hand side of \eqref{eq:res_one},} the drift term $\kperp\vdrift$ is being neglected. For improved accuracy, the drift term may be approximated as $(s/qR_0)\vpar$ with integer $s$ as in \citeref{Belikov2003POP} for very passing particles \addition{(a similar term would also appear due to the poloidal dependence of $\omegaci$)}. For the conditions in these simulations, the drift term is much smaller than $\kpar$ unless $s$ is quite large, $s \gtrsim 5$, which should be an inefficient resonant interaction. Due to these considerations, the analysis in this section will proceed with $s = 0$. The resonance condition can also be conveniently rewritten in terms of orbital frequencies as 
\begin{equation}
\label{eq:res_con}
\omega - n\avg{\omegator} - p\avg{\omegapol} = \lres\avg{\omegaci}
\end{equation}
\addition{Above, $p$ is in general an arbitrary integer, but equal to the negative poloidal mode number $(-m)$ when $s = 0$ as analyzed here.  For modes satisfying \eqref{eq:res_one} with $0 < \omega < \omegaci$ and $\vpar > 0$, counter propagation $(\kpar < 0$) implies $\lres > 0$, and co-propagation implies $\lres \leq 0$. In principle, particles could interact with the modes through $\abs{\lres} > 1$ resonances. However, this would require much larger Doppler shifts and particle velocities. A preference for $\abs{\lres} = 1$ is confirmed in the simulations. While the $\lres = 0$ resonance is present in the some of the simulations for the co-GAEs, it is usually subdominant to $\lres = 1$ (visible in \figref{fig:resonance}). Consequently, attention is restricted to the cases where $\lres = \pm 1$, which also leads to the correspondence $\lres = -\text{sign}\,\kpar$. Combining \eqref{eq:res_one}} with the presumed shear \Alfven dispersion, an expression can be written for the frequency of the excited mode as a function of the resonant $\vpar$ of the EP driving it unstable: 
\begin{equation}
\label{eq:res_vpar}
\omega = \frac{\avg{\omegaci}}{\lres + \avg{\vpar}/\va} \quad \text{for } \lres = \pm 1 = - \text{sign}\,\kpar
\end{equation}
Although $\vpar$ is not a constant of motion, it can be represented to lowest order in $\mu$ for each particle as 
\begin{equation}
\label{eq:vpar}
\vpar \approx v \sqrt{1 - \frac{\omegaci}{\omega_{ci0}}\lambda}
\end{equation}
\figref{fig:resonance} shows the parallel velocity (approximated by \eqref{eq:vpar}) of the EP with the largest weights, plotted against the frequency of the most unstable mode in each simulation. The relation between the mode frequency and parallel velocity of the most resonant particles generally adheres to \eqref{eq:res_vpar}, shown on the figures as the solid line. For co-GAEs, the condition is essentially obeyed, with some deviation due to a combination of drift term corrections and errors in the approximate expression for the resonant value of $\vpar$. In general, \eqref{eq:res_vpar} suggests that the frequency of the excited mode is inversely proportional to the parallel velocity of the resonant particles. While for co-GAEs the opposite trend is seen for fixed $n$ -- frequency increases with parallel velocity instead of decreases -- this is anticipated by the resonance condition. Since $\kpar \propto n$, the Doppler shift will increase with $\vpar$ at constant $n$. For cntr-GAEs, the mode frequencies still cluster near the curve representing \eqref{eq:res_vpar}, though there is substantial spread inherited from the deviations from the shear \Alfven dispersion due to ambiguous $\kpar$ as discussed in section \ref{sec:Structure}. 

\begin{figure}[tb]
\subfloat{\includegraphics[width = \halfwidth]{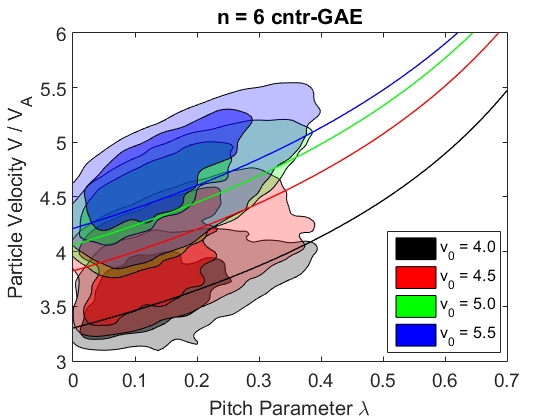}} \newline
\subfloat{\includegraphics[width = \halfwidth]{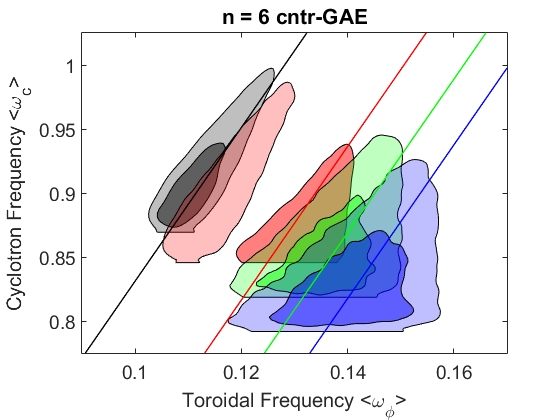}}
\caption{Resonant particles for $n = 6$ cntr-GAE excited by $\vinj = 4.0 - 5.5$. Top: Shaded contours show the location of the resonant particles in pitch-velocity $(\lambda,v)$ constant of motion space. Curves are contours of constant $\vpar$ determined by a $w$-weighted average of $\vpar$ over all resonant particles. Bottom: orbit-averaged toroidal and cyclotron frequencies of resonant particles. Solid lines show the resonance condition for each mode, averaging $\omegapol$ over all resonant particles and using the dominant $p$ in \eqref{eq:res_con} for each mode.}
\label{fig:res_conts}
\end{figure}

The mode frequency's sensitivity to the fast ions' location in phase space is reminiscent of energetic particle modes where the EPM frequency tracks typical particle orbit frequencies. Although the cyclotron and orbital frequencies are not constants of motion, a unique value of each can be calculated for each $\df$ particle as an orbit-averaged value. On \figref{fig:res_conts}, the shaded contours show the characteristic frequencies of the resonant particles in each simulation, where the resonant particles are defined as those with weights in the top $5\%$ at the end of the simulation. As the injection velocity increases, the resonant particles migrate to larger toroidal frequencies and smaller cyclotron frequencies. The lines imposed on the plot of toroidal vs cyclotron frequency represent the relation expected by the Doppler shifted cyclotron resonance in the form of \eqref{eq:res_con}. The resonant particles in each simulation cluster around these lines, showing that the frequency of the most unstable mode is being set by the location of the resonant particles in this phase space. In other words, the mode frequency adapts to the energetic particle attributes in order to satisfy the resonance condition. It is also helpful to examine where the resonant particles exist in the constant-of-motion space, $(v,\lambda,\pphi)$, which are the natural variables for the distribution function. This is shown in the top plot of \figref{fig:res_conts}. The resonant particles move towards higher energy as those regions become accessible with the larger injection velocity. For each distribution, a curve representing constant $\vpar$ is shown, with value determined by averaging over all resonant particles. Each shaded contour roughly tracks this line of constant $\vpar$, with value increasing with increasing $\vinj$. 

\begin{figure*}[tb]
\subfloat{\includegraphics[width = \halfwidth]{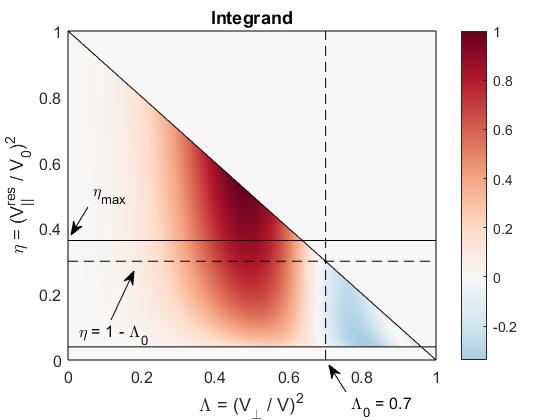}} 
\subfloat{\includegraphics[width = \halfwidth]{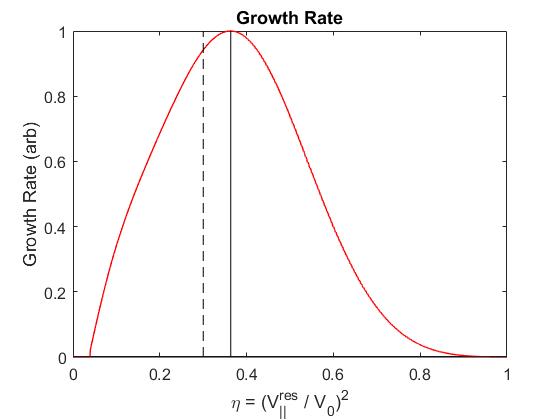}}
\caption{Left: integrand of growth rate integral, from \eqref{eq:gamma_full}, for a cntr-GAE with $\Lambda_0 = 0.7$. $\alpha = \kpar/\kperp = 0.5$ and $\vinj = 5.0$ are chosen as typical values. Vertical dashed line shows the central value $\Lambda_0 = 0.7$, horizontal dashed line shows the sufficient condition for net drive at $\Bres = \Epres/\E_0 = 1 - \Lambda_0$, and the solid line shows the value of $\Bres$ that maximizes the growth rate. Right: integral of left plot with respect to $\Lambda$, showing growth rate as a function of $\Bres$. Vertical lines match horizontal lines on left plot. Units are arbitrary in both since constants are excluded.}
\label{fig:integral}
\end{figure*}

Overall, \figref{fig:res_conts} demonstrates a clear linear relation between the energetic particle parameters and the frequency of the excited mode, a hallmark quality of energetic particle modes\cite{Todo2006POP}. This finding contradicts the conventional ``beam-driven MHD mode" paradigm where the energetic particles provide drive but otherwise do not affect the excited MHD mode. On the one hand, a resonant wave-particle interaction is necessary to drive the mode unstable, in which case it is natural that the frequency of the mode matches the combined orbital and cyclotron motion of the resonant particles. However, it is quite remarkable that the frequency of the mode is changing without clear changes in the mode structure. If this were a perturbative MHD mode, then one would expect that the changes in frequency would correspond to changes in mode structure, \ie poloidal or radial mode numbers. Alternatively, if only a single, specific eigenmode were being excited, then its frequency should not change as the energetic particle population does -- the mode would simply pick out the same resonant particles as $\vinj$ is increased. In view of these findings, this mode, formerly identified as a GAE from ideal MHD theory must be strongly altered by nonperturbative energetic particle effects, and thus could be considered as an energetic particle mode. This is different from the energetic particle modes commonly observed in experiments and discussed in the literature (fishbone, E-GAM, etc) typically have much lower frequencies, on the order of orbital frequencies\cite{Heidbrink2008POP}. To our knowledge, this is the first evidence of an EPM that is driven by a cyclotron resonance and with a frequency that can be an appreciable fraction of the cyclotron frequency.
\begin{figure*}[tb]
\subfloat{\includegraphics[width = \halfwidth]{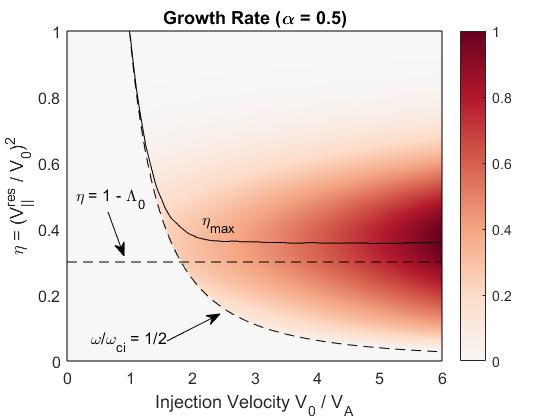}}
\subfloat{\includegraphics[width = \halfwidth]{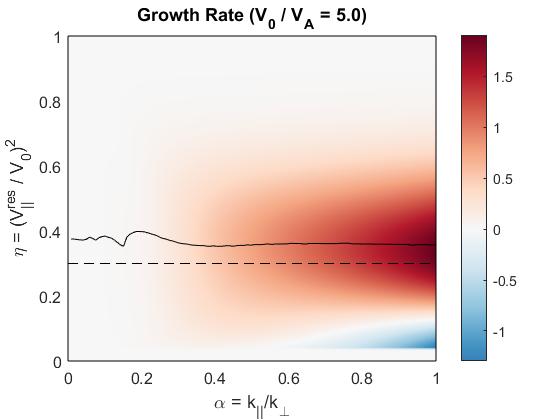}}
\caption{Left: Growth rate as a function of $\vinj$ and $\Bres = \Epres/\E_0$ for $\alpha = \kpar/\kperp = 0.5$. Dashed curve shows the $\omeganorm \leq 0.5$ boundary, where the integration is restricted to in order to satisfy $\omega \ll \omegaci$. Dashed horizontal line is the sufficient drive condition $\Bres = 1 - \Lambda_0$. The solid curve shows the value of $\Bres$ which maximizes the growth rate as a function of $\vinj$. 
Right: Growth rate as a function of $\alpha$ and $\Bres$ for $\vinj = 5.0$. Solid curve and dashed line have the same definition as to the left. Units are arbitrary in both since constants are excluded.}
\label{fig:gamma_scan}
\end{figure*}
\subsection{Relationship between injection and resonant velocities}
\label{sec:ResB}
The key takeaway is that if the resonant value of $\vpar$ is proportional to the injection velocity $\vb$, then the large frequency changes of these GAEs are qualitatively explained by the resonance condition. This is plausible based on the perturbative GAE growth rate expression derived by Gorelenkov \etal in \citeref{Gorelenkov2003NF}. The growth rate for the fast ion distribution defined in \eqref{eq:F0} is approximately proportional to 
\begin{equation}
\label{eq:gammaAppx}
\gamma \propto -\Bres^{3/2}\int_0^{1 - \Bres} d\Lambda 
\frac{\Lambda}{\left(1-\Lambda\right)^2}
\left(J_0(z) + J_2(z)\right)^2
(\Lambda - \Lambda_0) F_0
\end{equation}
This expression is derived in \appref{sec:GrowthRate}. Here $\Lambda \defined \lambda\omegaci/\omegacio = \Eperp/\E$, $z = \kperp\rho_\perp$, and $\Bres = \Epres/\E_0$ is the parallel energy of the resonant fast ions relative to the injection energy. All terms in the integrand are strictly nonnegative except for $-\left(\Lambda - \Lambda_0\right)$, which is positive for $\Lambda < \Lambda_0$. Therefore, for $\Bres > 1 - \Lambda_0$, the integral is strictly positive, which yields a \emph{sufficient} condition for net drive from the energetic particles. If the exact marginal stability were near this threshold, then it would explain why the simulation results imply that the resonant parallel velocity is proportional to the injection velocity. However, without reliable calculations of the damping of the mode due to interaction with the bulk plasma -- most notably continuum damping -- to compare with, this sufficient condition can lend intuition but not a definitive explanation of why $\vpar^{res}$ is seemingly proportional to $v_0$, since the marginal stability condition could shift substantially depending on the magnitude of the continuum damping.  

To complement the preceding argument regarding the condition for marginal drive from the fast ions, the unsimplified growth rate derived in the appendix in \eqref{eq:gamma_full} can be evaluated numerically to determine how the maximum growth rate depends on its three independent parameters $\vinj$, $\Bres$, and $\alpha = \kpar/\kperp$ for a cntr-GAE $(\lres = +1)$ with $\linj = 0.7$. The sum of Bessel functions embedded in $\Jlm(z)$ (see \eqref{eq:Jlg}) is the main obstacle to gaining intuition about the growth rate's dependencies by inspection or calculus. The parameter $\alpha$ enters through this Bessel term, since its argument can be rewritten as
\begin{equation}
\label{eq:ztransform} 
z = \kperp\rho_\perp = \frac{\kperp\vperp}{\omegaci} = 
\frac{\kperp}{\kpar}\frac{\kpar\va}{\omegaci}\frac{\vperp}{\va} = 
\frac{1}{\alpha}\frac{\omega}{\omegaci}\frac{v_0}{\va}\sqrt{\frac{\Bres\Lambda}{1-\Lambda}}
\end{equation}

Above we have used $\omega \approx \kpar\va$, which is an approximation made in the calculation which arrived at \eqref{eq:gammaAppx}. Note also that $\omega$ is not an independent parameter due to \eqref{eq:res_vpar}. Moreover, this calculation is only valid for $\omega \ll \omegaci$, so the integration will be restricted to $\omega/\omegaci < 0.5$. A common tokamak approximation is $\kperp \gg \kpar$, or equivalently $\alpha \ll 1$. The \HYM simulations reveal that while $\kperp \gtrsim \kpar$, it is not significantly greater for these modes in the compact NSTX geometry. Fourier transforms of the numerical mode structure yield characteristic values of $\alpha = 0.3 - 1$. 

The integrand with $\vinj = 5.0, \alpha = 0.5, \Lambda_0 = 0.7$ is shown in the left of \figref{fig:integral}, revealing complicated dependence on both the integration variable $\Lambda$ and the parameter $\Bres$. Generally, decreasing $\alpha$ makes the details of the integrand even more intricate, as more zeros of $\Jlm(z)$ become contained within the integration region. Visualized this way it is clear why the sufficient condition for net drive from the energetic particles exists: at sufficiently large $\Bres$, the upper integration bound excludes the regions of velocity phase space which damp the wave. The cutoff at very small $\Bres$ is imposed due to the condition $\omega/\omegaci < 0.5$, which ensures that $\omega \ll \omegaci$. The right plot in \figref{fig:integral} shows the growth rate's dependence on $\Bres$ for these specific values of $\vinj$ and $\alpha$, demonstrating a local maximum exceeding the sufficient threshold for net drive at $\Bres = 1 - \Lambda_0$ (dashed line). This optimal value of $\Bres$ is also marked on the left plot with the solid line near $\Bres = 0.36$. 

Numerical integration can be performed over a range of values of $\vinj$, $\Bres$, and $\alpha$ in order to determine if the growth rate prefers changing the frequency of the mode as $\vinj$ is varied, which would explain the simulation results. These scans are shown in \figref{fig:gamma_scan}. For $\vinj \gtrsim 2.5$ and $\alpha \gtrsim 0.4$, there is a clear preference for $\Bres \approx 0.36$ in order to maximize the growth rate, as the optimal value of $\Bres$ is within $1\%$ of this value in this range of parameters, which also encompasses the properties of the simulated modes. This calculation implies that the energetic particle drive is maximized for a mode resonantly excited by a subpopulation of fast ions with parallel velocity at a specific fraction of the injection velocity, explaining the connection between the injection velocity and resonant parallel velocity. Then, the frequency dependence due to the resonance condition becomes 

\begin{equation}
\label{eq:res_exp}
\omega = \lres\omegaci + \kpar v_0\sqrt{\Bres} = \lres\left[\omegaci - \abs{\kpar} v_0\sqrt{\Bres}\right]
\end{equation}

Above we have used the fact that $\lres = - \text{sign}\,\kpar$. In the case of cntr-GAEs $(\lres = +1)$, the Doppler shift is less than the cyclotron frequency, and so the preferred mode frequency decreases linearly as a function of $v_0$. Conversely, co-GAEs excited by the $\lres = -1$ resonance have a Doppler shift exceeding the cyclotron frequency, so the frequency of the most unstable mode will increase linearly with increasing $v_0$. While this result reproduces the frequency trend of the most unstable modes from the simulations, the calculation is limited by not including the sources of bulk plasma damping. It is fair to assume that the thermal damping will affect each mode similarly, and hence, the maximum growth rate argument could remain valid. However, the amount of continuum damping each mode is subject to could vary substantially depending on quantitative details of the mode structure and differences in the self-consistent equilibria generated by fast ion populations of different injection velocities. Simplified analytic calculations have been performed in order to understand the numerical results, and they do not include the effects of continuum damping. Nonetheless, the presence of this frequency dependence both in simulations with signs of coupling to the continuum (via the appearance of short scale structures near the ideal \Alfven resonance location) as well as in those where they are absent indicates that the impact of continuum damping may not be crucial to developing a qualitative understanding of this phenomenon. The determination of the most unstable mode based on maximizing drive from the fast ions may be suitable to describe the robust numerical results. 
\section{Summary and Discussion}
\label{sec:Discussion} 
Hybrid simulations have been conducted to study how the properties of high frequency shear \Alfven eigenmodes depend on parameters of the energetic particle distribution in NSTX-like low aspect ratio conditions. In simulations that solve for the equilibrium with self-consistent inclusion of energetic particle effects, it is found that the frequency of the most unstable GAE changes significantly with the energetic particle parameters. The frequency changes most significantly with the normalized injection velocity $\vinj$, which shows a clear linear relation. With increasing injection velocity, counter-propagating modes have a decrease in frequency, while co-propagating modes increase in frequency. The linear dependence and sign of the change are consistent with the Doppler-shifted cyclotron resonance condition. 

However, there are no clear concurrent changes in mode structure that would indicate that these frequencies correspond to distinct eigenmodes, especially for the co-GAEs. Moreover, the frequencies change continuously as a function of the injection velocity, not in a discrete stair-stepping pattern one would expect if different discrete eigenmodes were being excited. In contrast, the frequencies of  compressional modes excited in the simulations are largely unaffected by the fast ions, and modes with distinct frequencies have different poloidal mode numbers. 

At fixed injection energy, the frequency of both co- and counter-propagating modes decrease as the normalized EP density $\nb$ is increased, though the frequency change is an order of magnitude less than that caused by changing the injection energy. Although there was some difficulty in determining a reliable value of $\kpar$ for these modes due to low aspect ratio and poorly defined $m$ numbers, the modes do roughly obey the shear \Alfven dispersion relation $\omega \approx \left[\kpar(r)\va(r)\right]_{r=r_0}$, evaluated at the mode location, to within $10 - 20\%$. Lastly, the substantial changes in frequency persist even when the energetic particles are ignored in the equilibrium solver, implying that the change in frequency directly due to changes in the energetic particle population is much larger than the indirect change in frequency due to changes in the equilibrium from fast particle contributions. 

Put together, these results call into question the description of these modes as the global \Alfven eigenmodes described by ideal MHD theory. Since GAEs are shear \Alfven MHD modes, in order to be weakly damped they must have frequencies just below a minimum of the \Alfven continuum. Large frequency shifts with changing beam parameters can displace the modes from being localized near these extrema, and lead them to intersect the continuum where they woudl be expected to suffer strong damping. The energetic particles are clearly exerting a nonperturbative effect on the modes since the eigenfrequency is changing without clear corresponding changes in the mode structure that would indicate excitation of a different eigenmode. Instead, these results could be interpreted as defining a high frequency energetic particle mode, regarded here as an energetic-particle-modified global \Alfven eigenmode (\EGAE). For excitation, the mode must be resonant with a sub-population of energetic particles with a specific value of $\vpar$. As the injection velocity is increased, new values of $\vpar$ become accessible. It was shown that the drive from the fast ions is maximized for a resonant parallel velocity at a specific fraction of the injection velocity, given the same degree of anisotropy. As the resonant value of $\vpar$ changes, both $\omega$ and $\kpar$ must also change according to the resonance condition and the approximate dispersion. An energetic particle mode defined by a continuum of $\kpar$ values to choose from as the injection velocity is varied is consistent with these findings. This is unusual since energetic particle modes typically have much lower frequencies which track the characteristic energetic particle orbital frequency\cite{Todo2006POP}. In contrast, the modes excited in these simulations can be an appreciable fraction of the cyclotron frequency, $\omega \approx 0.1 - 0.5\omegaci$ for the range of toroidal harmonics $\abs{n} = 4 - 16$, and have frequencies which track a combination of the energetic particle orbital and cyclotron frequencies. 

There have been previous studies showing an MHD mode's eigenfrequency changing in proportion to energetic particle velocity. One is the so-called ``resonant toroidicity-induced \Alfven eigenmode" (RTAE), which is characterized by the mode frequency decreasing in order to remain in resonance with fast particles as $T_{EP}/T_i$ decreases\cite{Cheng1995NF}. Cheng \etal remark that this trend can lead the RTAE to have a frequency much below the characteristic TAE gap frequency that it is associated with, just as the GAEs in these simulation results can be significantly displaced from the minimum in the \Alfven continuum. In addition, previous hybrid gyrokinetic simulations have demonstrated a transition from TAE to a lower frequency kinetic ballooning modes (KBM) as the maximum energetic particle energy is increased\cite{Santoro1996PP}. During this transition, the frequency of the KBM changes in proportion to the energetic particle velocity, similar to the results presented here. 

Although the exact dispersion of the \EGAE has not yet been determined, it is clear that it is fundamentally affected by the energetic particles nonperturbatively, leading to a departure from its previous perturbative MHD description. In addition to the interest to basic plasma physics of the discovery of a high frequency energetic particle mode with frequencies tracking the combined orbital and cyclotron motion, there are also potential implications for NSTX-U which should be explored in the future. The simulations presented here show that the nonperturbative regime for these modes was routinely accessed in NSTX operating conditions. The basic picture of an energetic beam driving an MHD mode of the thermal plasma without modifying its attributes breaks down in conditions where $\Jb$ is comparable to $\Jp$. Even with the nominal factor of two increase in toroidal field in NSTX-U which will tend to decrease $\vinj$, these modes may still be unstable due to the increase in beam power\cite{Gerhardt2012NF}, though early operations indicate they can be suppressed with the addition of off-axis injection\cite{Fredrickson2017PRL}. 

NSTX experiments have established a robust link between sub-cyclotron \Alfven modes and anomalous electron temperature flattening\cite{Stutman2009PRL,Ren2017NF}. Both of the existing theoretical mechanisms proposed to explain how \Alfvenic modes could generate this anomalous heat diffusivity have previously assumed that they are accurately described as perturbative ideal MHD GAEs\cite{Gorelenkov2004POP,Kolesnichenko2010PRL,Gorelenkov2010NF}. Since it has now been shown that there can be quite substantial nonperturbative corrections to this description, the polarization and mode structure of these modes may be quite different from those assumed by these previous analyses. In particular, Gorelenkov \etal investigated how several overlapping GAEs could collectively stochasticize electron orbits and enhance the radial diffusion. Nonperturbative modifications of the mode characteristics could alter the thresholds in number of overlapping modes and mode amplitudes required to generate the level of diffusion necessary to explain the experimental observations. While compressional modes have received more attention for their potential to channel energy away from the core to the edge through mode conversion to kinetic \Alfven waves\cite{Belova2015PRL,Belova2017POP}, GAEs also couple to KAWs in principle\cite{Kolesnichenko2010PRL,Kolesnichenko2010NF} and may also contribute. At least in the case of GAE-KAW mode conversion, the simulation results presented here suggest that nonperturbative inclusion of the energetic particles should be further explored for a more accurate description of that coupling in application to energy channeling in fusion conditions. Examining the impact of these corrections on previous quantitative predictions of anomalous electron heat transport will be the subject of future work. 

Prospects for future experimental verification of the EP-GAE are promising, as its defining characteristics should be observable in suitably designed experiments on NSTX-U. Analysis without such dedicated experiments may prove challenging since it is necessary to separate the changes in mode frequency due to the change in beam energy (the nonperturbative effect) from the changes in the equilibrium (MHD effect). The preferred approach would be to reproduce a discharge multiple times with different beam voltages for each shot so that the time evolution of the equilibrium profiles can be factored out of the observed change in frequency, such as the experiments conducted in \citeref{Fredrickson2002POP}. Measurement of the change in frequency due to this effect could be further complicated by chirping, which sometimes occurs for the high frequency \Alfvenic modes in NSTX. Fortunately, existing analysis shows that this usually takes the form of symmetric chirping (as opposed to monotonic frequency sweeping) about the linear mode frequency\cite{Fredrickson2006POP}. In this case, the frequency dependence on $\vinj$ should still be detectable. In addition to the signature change in frequency in proportion to the injection velocity, the gradual shift of the counter-propagating mode further towards the low field side with increasing beam energy as discussed in \secref{sec:Structure} may be observable with reflectometer measurements\cite{Crocker2013PPCF,Crocker2013NF}. 
\section{Acknowledgements}
\label{sec:Acknowledgements}
The simulations reported here were performed with computing resources at the National Energy Research Scientific Computing Center (NERSC). The data required to generate the figures in this paper are archived in the NSTX-U Data Repository ARK at the following address: \url{http://arks.princeton.edu/ark:/88435/dsp01s1784p39h}. This research was supported by the U.S. Department of Energy (NSTX contract \# DE-AC02-09CH11466).
\appendix
\section{GAE Growth Rate} 
\label{sec:GrowthRate}
The GAE growth rate for $\omeganorm \ll 1$ is calculated perturbatively for the fast ion distribution function used in these simulations. Beginning with Eq. 16 of \citeref{Gorelenkov2003NF} and ignoring coefficients, the growth rate is proportional to
\begin{equation}
\gamma \propto \int d\E d\E_\perp I^2\delta\left(\theta - \theta_{res}\right)
\vec{G}_{\lres}^{\prime *}\dot \vec{E}^{*}\vec{G}_{\lres}\dot\vec{E} 
\left[\frac{\partial}{\partial\E} + \frac{\lres\omegaci}{\omega}\frac{\partial}{\partial\E_\perp}\right]F_0
\end{equation}
The delta function enforces the resonance condition by evaluating the integrand at the resonant locations along the particle's trajectory. Additionally, $I^2$ is the resonance factor defined in Eq. 47 of \citeref{Gorelenkov1995POP}, and represents the time duration of the wave-particle interaction in one pass through the resonance layer. Combination of Eq. 16 and 17 from \citeref{Gorelenkov2003NF} implies the simplified relation $I^2\vpar\delta(\theta - \theta_{res})/8\pi q R = \delta(\omega - \kpar\vpar - \lres\omegaci)$, which in concert with the delta function identity $\delta(f(x)) = \delta(x-x_0) / \abs{f'(x_0)}$ yields the transformation $I^2\delta(\theta - \theta_{res}) \propto \delta(\Epar - \Epres)/\abs{\kpar}$, ignoring constants. 

Moreover, $\vec{G}_\lres = \vperp(-iJ_\lres^\prime(z), J_\lres(z)/z), \vec{G}_\lres^\prime = \vperp(-iJ_\lres^\prime(z),\lres J_\lres(z)/z)$, where $J_\lres$ is the Bessel function of the first kind of order $\lres$ with argument $z = \kperp\rho_\perp = \kperp\vperp/\omegaci$. Also introduce $\alpha\defined \kpar/\kperp$. Then the GAE polarization is $E_1 = i (\alpha^2\omega/\omegaci) E_2 \ll E_2$, and the coordinates are defined by $\hat{\vec{2}} = \vec{\kperp}/\kperp$ and $\hat{\vec{1}}\times\hat{\vec{2}} = \vec{B_0}/B_0$. 

Define $\Jlm(z)$ via $\vec{G}_{\lres}^{\prime *} \dot \vec{E}^{*}\vec{G}_{\lres}\dot\vec{E} = \abs{E_2}^2\vperp^2 \Jlm(z)$. The $\lres$ subscript is the resonant cyclotron coefficient, and the $m$ superscript is either $G$ for GAEs or $C$ for CAEs. The full expression for $\Jlg$ is

\begin{equation}
\label{eq:Jlg} 
\Jlg = \frac{\lres J_\lres^2}{z^2} + (1 + \lres)\oba\frac{J_\lres J_\lres^\prime}{z} + \left(\oba\right)^2J_\lres^{\prime 2}
\end{equation}

To leading order in $\alpha^2\omega/\omegaci \ll 1$, $\Jlg \approx \abs{E_2}^2\vperp^2\lres(J_\ell/z)^2 = \vperp^2\left(J_{\lres-1} + J_{\lres+1}\right)^2/4\lres$ for $\lres \neq 0$, which is the dominant resonance for the modes studied here. For $\lres = \pm 1$, $J_{\lres-1} + J_{\lres+1} = J_0 + J_2$ since $J_{-\nu} = (-1)^{\nu}J_{\nu}$. However, all terms in $\Jlg$ will be kept for numerical integration, since they can be important when $\omega/\omegaci$ and $\alpha$ are small but not trivially so.  

Defining the operator in brackets as $\fgrad \defined \frac{\partial}{\partial \E} + \lres(\omegaci/\omega)\frac{\partial}{\partial \Eperp}$, it can be rewritten into derivatives with respect to $\E$ and $\Lambda \defined \lambda \omegacio/\omegaci = \Eperp/\E$ for convenience. 
\begin{align}
\label{eq:PiF0}
\begin{split}
\fgrad F_0 &= \frac{1}{\E}\left[\E\frac{\partial}{\partial \E} + 
\left(\frac{\lres\omegaci}{\omega} - \Lambda\right)\frac{\partial}{\partial\Lambda}\right]F_0 \\
&= -\frac{2}{\E}\left[\frac{3}{4}\frac{1}{1 + \left(\E_0/4\E\right)^{3/2}} 
+ \left(\frac{\lres\omegaci}{\omega} - \Lambda\right)\left(\frac{\Lambda-\Lambda_0}{\dl^2}\right)\right]F_0
\end{split}
\end{align}

Interestingly, the chosen equilibrium fast ion distribution is an eigenfunction of this operator. 
Note that for $\omega/\omegaci \ll 1$, $(\lres\omegaci/\omega - \Lambda) \approx \lres\omegaci/\omega$. 
Furthermore, the first term in \eqref{eq:PiF0} is at most $3/4$, which is much less than the second term, except in the small region where $(\Lambda-\Lambda_0) \lesssim 3\dl^2\omega/4\omegacio < 0.05$ even for $\omega/\omegacio = 1/2$. Numerical integration can demonstrate that this region does not contribute to the integral significantly. Thus $\fgrad F_0$ can be approximated as
\begin{equation}
\fgrad F_0 \approx -\frac{2\lres\omegaci}{\dl^2\omega\E}\left(\Lambda - \Lambda_0\right)F_0
\end{equation}

Lastly, change variables of integration from $(\E,\Eperp)$ to $(\Epar,\Lambda)$, which has the Jacobian $d\E d\Eperp = [\Epar / (1-\Lambda)^2] d\Epar d\Lambda$. After performing the integration over $\Epar$ with the delta function to enforce the resonance condition and introducing $\Bres = \Epres/\E_0$, the growth rate is approximately proportional to
\begin{equation}
\label{eq:gamma}
\gamma \propto -\Bres^{3/2}\int_0^{1 - \Bres} d\Lambda 
\frac{\Lambda}{\left(1-\Lambda\right)^2}\lres\Jlg
(\Lambda - \Lambda_0) F_0
\end{equation}
The integration bound results from $\Epar = \E\left(1 - \Lambda\right) < \E_0\left(1 - \Lambda\right)$. Since the integrand is nonnegative for $\Lambda < \Lambda_0$ (note $\lres\Jlg$ is positive except where it is very small), the integral is strictly positive for $\Epres/\E_0 > 1 - \Lambda_0$. This represents a sufficient condition for net drive of the wave due to the contribution from the fast ions. Without these simplifications, and denoting $A(\Lambda_0,\dl)$ as a complicated normalization function, the growth rate for $\lres\neq 0$ is proportional to
\begin{multline}
\label{eq:gamma_full}
\gamma \propto -\frac{\Bres^{3/2}}{\abs{\omega -\lres\omegaci}}
\int_0^{1-\Bres} d\Lambda \frac{\Lambda}{\left(1-\Lambda\right)^2}\Jlm(z) \times  \\ 
\left[\frac{3}{4}\frac{1}{1 + \left(\frac{1-\Lambda}{4\Bres}\right)^{3/2}} 
+ \left(\frac{\lres\omegaci}{\omega} - \Lambda\right)\left(\frac{\Lambda - \Lambda_0}{\Delta\lambda^2}\right)\right] \times \\ 
\frac{A(\Lambda_0,\dl)\nbo v_0}{\left(\frac{\beta}{1-\Lambda}\right)^{3/2} + \frac{1}{8}}
\exp{\left(-\frac{\left(\Lambda - \Lambda_0\right)^2}{\dl^2}\right)}
\end{multline} 

\bibliography{GAE_shift_bib}
\end{document}